# Geovisual Analytics and Interactive Machine Learning for Situational Awareness

Morteza Karimzadeh [a, *], Luke S. Snyder [b], David S. Ebert [b]

[a] *Department of Geography, University of Colorado Boulder (formerly at Purdue University), karimzadeh@colorado.edu*
[b] *School of Electrical and Computer Engineering, Purdue University, snyde238@purdue.edu, ebertd@purdue.edu*

* Corresponding author

**Abstract**

The first responder community has traditionally relied on calls from the public, officially provided geographic information, and maps for coordinating actions on the ground. The ubiquity of social media platforms created an opportunity for near real-time sensing of a situation (e.g., an unfolding weather event or crisis) through volunteered geographic information. In this article, we provide an overview of the design process and features of Social Media Analytics Reporting Toolkit (SMART), a visual analytics platform developed at Purdue University for providing first responders with real-time situational awareness. We attribute its successful adoption by many first responders to its user-centered design, interactive (geo)visualizations, and interactive machine learning, thus giving users control over analysis.

**Introduction**

Social media has created an opportunity for researchers and practitioners to harvest volunteered geographic information for various purposes, including sentiment analysis, mobility and movement studies, polling, and market analysis. The real-time nature of social media makes it an excellent source for situational awareness and crisis management. For instance, several emergency operation centers' staff members have told us that they receive useful information about a situation on the ground at least 10–15 minutes prior to receiving phone calls from the public. However, the data obtained through social media is big, with relevant information overwhelmed by noise. It is crucial to harness and summarize the sea of data according to first responders' needs, which may vary depending on the situation.

**Social Media Analytics and Reporting Toolkit**

Social Media Analytics Reporting Toolkit (SMART) is an interactive, web-based visual analytics system targeted for first responders who may not have computational expertise (Zhang et al. 2017; 2014; Snyder, Lin, et al. 2019). It leverages many integrated algorithms and technologies for scalable, real-time, and interactive social media (currently, only public Tweets) analysis and visualization. SMART allows users to filter Tweets based on their topic through the inclusion and exclusion of keywords. Further, it provides interactive, human-in-the-loop machine learning of relevant posts to each analytical scenario. After Tweets are filtered, users can use multiple interactive visualizations to view topics aggregated in time (e.g., the ThemeRiver visualization [Havre et al. 2002]), over space (e.g., spatial topic modeling, spatial content lenses, and spatial topic lenses), by topics (e.g., word clouds and topic modeling), or by combinations of these dimensions. These different visualizations are linked, allowing users to further inspect and drill down to certain topics, spatial extents, or time periods at multiple scales of aggregation. SMART combines advanced statistical modeling, text analytics, machine learning, and novel anomaly detection techniques augmented by human expertise so that users can identify trending and abnormal topics on social media. Figure 1 provides an overview of SMART's user interface.

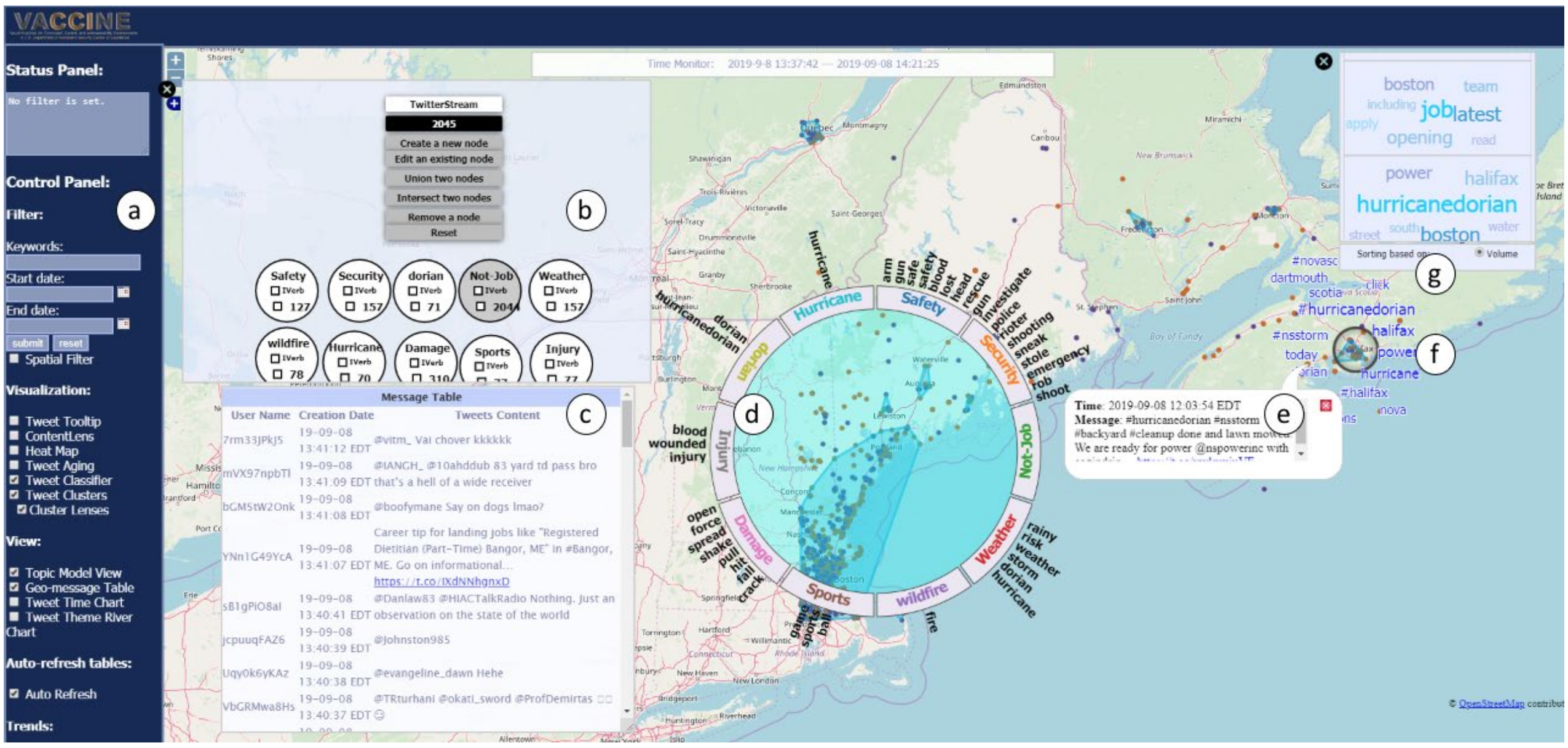

Figure 1. SMART's web interface and interactive visualizations. The control panel (a) allows users to activate different components. The Tweet classifier (b) allows users to construct semantic keyword filters to identify relevant information. The geomessage table (c) lists Tweets for direct content analysis. The (d) cluster lens provides a radial layout of keywords derived from each Tweet classifier and used within a spatial region. The Tweet tool tip (e) allows users to directly view a Tweet's content from the map. The content lens (f) uses topic modeling to highlight the most frequently used words within an area. The word cloud view (g) uses topic modeling to extract trending topics and associated keywords.

Giving users maximum control of system content, behavior, and visualization has been central to SMART's success. We fulfilled this vision through customizable topical filters, interactive maps and visualizations, and interactive machine learning with minimal user feedback. Even though keyword filtering provides basic functionality for users to narrow the stream of social media posts down to the topics that they are interested in, the amount of noise and irrelevant information may still be overwhelming. Furthermore, user intentions and situational requirements vary from one event to another. For instance, a particular topic such as traffic may be of interest during an event such as a wildfire adjacent to highway networks but not during a wildfire that occurs in the wilderness. Therefore, SMART provides interactive learning functionality (based on a shallow neural network) in which users can mark posts as being relevant or not relevant. According to our evaluations, the system learns from the user quickly, with more than ~50 percent accuracy after 50 labels have been provided by the user, and reaching accuracies up to ~80 percent in identifying relevant posts after ~200 labels have been provided by the user (Snyder, Lin, et al. 2019). However, combined with interactive sorting and zooming, even fewer clicks by the user sufficiently tailors the system and the visualized content to user analytical needs. Figure 2 shows the results of interacting with the system after only 20 clicks by the user to train the integrated neural network model and sorting the Tweets based on relevance to the topic—in this case, "weather"—chosen by a user. This functionality enables first responders to quickly identify relevant information from social media that also includes large amounts of noisy and irrelevant data.

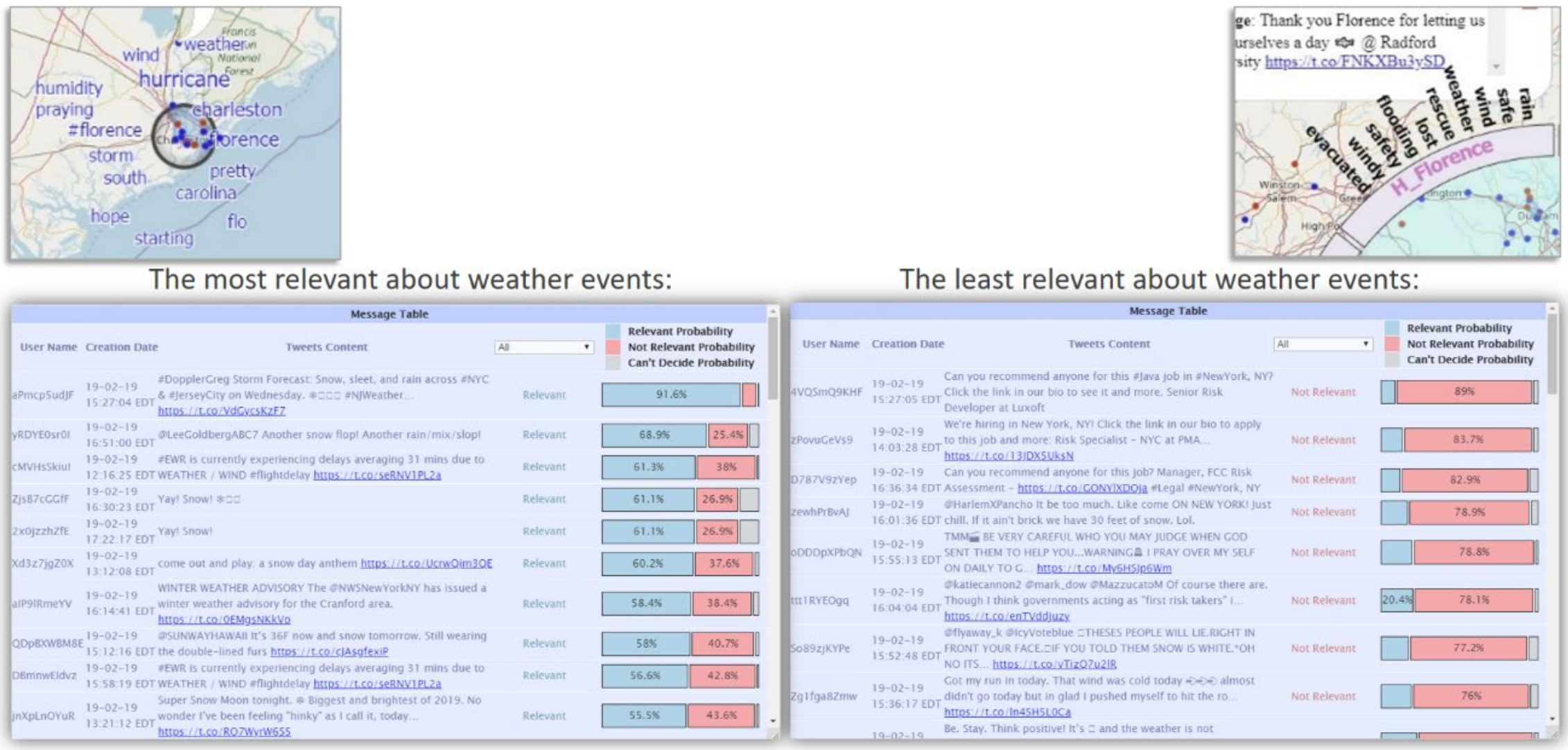

Figure 2. Interactive machine learning for classifying social media posts as relevant/irrelevant to a weather event. By training a classifier, the analyst can see relevant Tweets on interactive maps. The table on the left shows the Tweets sorted by descending relevance to the topic of weather, and the table on the right shows Tweets sorted by ascending relevance. These are achieved after only 20 user clicks, which is a significant improvement over mere keyword-based filtering.

**User-Centered Design and the Consequent Adoption of SMART**

SMART was developed at Purdue University with the goal of supporting shared situational awareness so that decision-makers at every level have access to the same crowdsourced information from public data feeds. Central to SMART's design and development have been the continuous stakeholder engagement, requirement analysis, demos, interviews, feedback sessions, agile development, and refinements (Snyder, Karimzadeh, et al. 2019). Through the adoption, modification, improvement, and extension of state-of-the-art approaches (Havre et al. 2002, Karimzadeh et al. 2019, Bosch et al. 2011, Thom et al. 2012), SMART's development has primarily been focused on meeting the user requirements rather than exploring academic curiosity. However, the real-world use of SMART has enabled significant research accomplishments at the forefront of computer science and (geographic) information science, reported in various conference and journal articles (Chae et al. 2014; Zhang et al. 2017; Snyder, Lin, et al. 2019; Snyder, Karimzadeh, et al. 2019).

SMART has been successfully used in many events since its first utilization at the Boy Scouts of America's 2013 National Scout Jamboree. Local emergency response departments, campus police departments, nongovernmental organizations, and the US Coast Guard (USCG) have used SMART extensively since then, with ~20 organizations actively using it for special events (e.g., festivals) or natural events (e.g., hurricanes, floods). For instance, during the 2017 hurricane season, the USCG used SMART to find people in distress and to target search and rescue efforts. SMART was also used for the 2016 Republican National Convention in Cleveland, Ohio; for the 2017 Presidential Inauguration; for the 2018 and 2019 State of the Union Addresses in Washington, DC; by state-level officials for the Thunder Over Louisville event; and for the Cincinnati Riverfest and several state fairs. In each case, Purdue University researchers trained the prospective end users in a one-hour webinar; sought feedback after use; and made necessary modifications to the system, which have now occurred over the course of several years.

Purdue University has licensed SMART to a startup company—supported by Purdue Foundry—for commercialization. This company will continue to support the existing end-user groups and enhance the scalability and usability of SMART for more users in the future.

**Conclusion and Future Directions**

Social media has proved useful in facilitating situational awareness, provided that computational methods and human expertise are combined in efficient ways. We achieved this through iterative, user-centered design with constant stakeholder engagement. Central to our approach was the full utilization of previous research while keeping users and interactivity in mind. We designed (or adopted), optimized, and evaluated our computational methods in interactive settings to ensure full user control over the system and its content. Users train the integrated machine learning models based on their needs; therein lies one of the primary reasons for SMART's successful adoption.

Moving forward, research and development should focus on both the technical and ethical aspects of privacy and geolocation to sustain situational awareness for social good while preventing adverse use or breaking society's trust. Many users are hesitant to share their geolocation directly but still provide approximate geolocation in their profiles, providing an opportunity for situational awareness with reduced chances of unwittingly flouting user privacy.

**Acknowledgement**

This material is based upon work funded by the US Department of Homeland Security (DHS) under DHS Award No. 2009-ST-061-CI0003, DHS Cooperative Agreement No. 2014-ST-061-ML0001, and DHS Science and Technology Directorate Award No. 70RSAT18CB0000004.